\documentclass[reprint,amsmath,amssymb,aps,]{revtex4-1}
\usepackage{graphicx}
\usepackage{dcolumn}
\usepackage{bm}
\usepackage{amsmath}
\usepackage{epstopdf}

\begin{document}
\preprint{APS/123-QED}
 
\title{Mind the gap but also the spin: why the Heyd-Scuseria-Ernzerhof hybrid functional description of VO$_2$ phases is not correct}
\author{Ricardo Grau-Crespo}
\email{r.grau-crespo@ucl.ac.uk}
\address{ Department of Chemistry, University College London, 20 Gordon Street, London WC1H 0AJ, UK}
\author{Hao Wang}     
\author{Udo Schwingenschl\"ogl}
\email{udo.schwingenschlogl@kaust.edu.sa}
\address{KAUST, PSE Division, Thuwal 23955-6900, Kingdom of Saudi Arabia}

\date{\today}

\begin{abstract}
In contrast with recent claims that the Heyd-Scuseria-Ernzerhof (HSE) screened hybrid functional can provide a good description of the electronic and magnetic structure of VO$_2$ phases [V. Eyert, \emph{Phys. Rev. Lett.} 107, 016401 (2011)], we show here that the HSE lowest-energy solutions for both the low-temperature monoclinic (M1) phase and the high-temperature rutile (R) phase, which are obtained upon inclusion of spin polarization, are at odds with experimental observations. For the M1 phase the groundstate is (but should not be) magnetic, while the groundstate of the R phase, which is also spin-polarized, is not (but should be) metallic. The energy difference between the low-temperature and high-temperature phases is also in strong discrepancy with the experimental latent heat. 
\end{abstract}

\maketitle

The screened hybrid functional approach of Heyd, Scuseria and Ernzerhof (HSE) has accumulated significant success in the description of structural and electronic properties of molecules and solids at a moderate computational cost \cite{heyd03, heyd06, heyd04}. It has been recently argued that density functional theory (DFT) calculations based on this functional are well capable of describing the electronic and magnetic properties of the metallic and insulating phases of vanadium dioxide VO$_2$ \cite{eyert11}. In particular, it was shown there that HSE calculations are capable of producing the expected band gap in the electronic structure of the monoclinic phase, in contrast with other DFT approximations \cite{eyert02}.   This is exciting progress, because modelling the VO$_2$ phase transition with relatively inexpensive DFT methods (the alternative is many-body GW calculations or dynamic mean-field theory \cite{biermann05, gatti07, weber12}) opens the door to a more active role of ab initio design in the development of applications such as ``smart" thermochromic windows \cite{parkin08}.
 
When heated to $\sim$340 K, pure VO$_2$ exhibits a transition from a monoclinic (M1) semiconductor phase to a tetragonal, rutile-like, metallic phase (R) \cite{morin59}. The transition is first-order \cite{corr10}, and a latent heat of  44 meV per VO$_2$ formula unit has been measured by calorimetric methods \cite{pintchovski78}. Therefore it can be expected that accurate calculations yield a band gap for the M1 phase but not for the R phase, and that the calculated total energy per formula unit of the R phase is higher than that of the M1 phase. In order to compare the electronic structures and energies of the two phases we have performed HSE calculations using the planewave DFT program VASP \cite{kresse96}, first using non-spin polarized calculations as in \cite{eyert11}, and then using spin-polarized calculations in different magnetic configurations. Experimentally determined crystal structures were used in the calculations \cite{rogers93}, including four and two formula units for the M1 and R phases, respectively, without geometry relaxation.  In order to ensure a reliable energy comparison between phases, precision parameters were chosen to achieve a convergence of 1 meV/atom in total energy. This required k-point grids of 5x5x5 and 6x6x9 for the M1 and R unit cells, respectively (the same grids were used for the non-local exchange contributions). The energy cutoff for the planewave basis set was 400 eV, and the interaction between the core (up to 3p for V and up to 1s for O) and valence electrons was described with the projected augmented wave (PAW) method \cite{blochl94}. The standard settings were used for the screened hybrid functional, i.e. 25\% of Hartree-Fock exchange was mixed in, with a screening parameter $\mu=0.207 \AA^{-1}$, and the local contributions were calculated with the Perdew-Burke-Ernzerhof (PBE) functional \cite{perdew96}. 

\begin{figure}
\includegraphics[width=7cm]{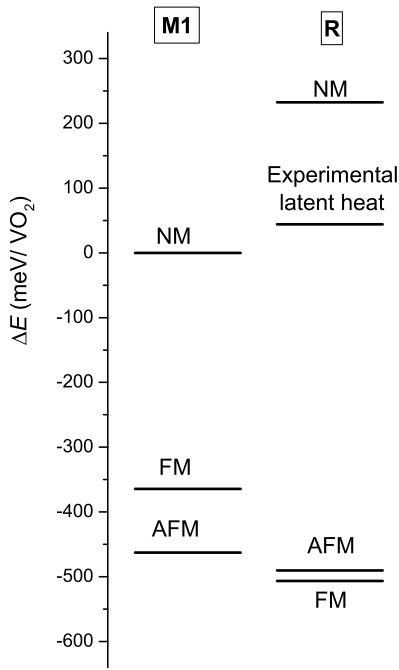}
\caption{\label{fig:one} Energies per VO$_2$ formula unit of the monoclinic (M1) and rutile (R) phases of VO$_2$ as obtained with the HSE functional in nonmagnetic (NM), ferromagnetic (FM) and antiferromagnetic (AFM) configurations. All energies are given with respect to the NM solution for the M1 phase, which should be (but is not for HSE) the global groundstate at zero temperature. The experimental latent heat is taken from reference \cite{pintchovski78}.}
\end{figure}

The results for the non-magnetic (NM) calculations  in Fig. \ref{fig:one} show that the energy of the R phase is indeed higher than that of the M1 phase. However, the calculated energy difference ($E$[R]-$E$[M1]=232 meV per VO$_2$ formula unit) between the two phases is too large compared with the experimental latent heat. Our estimation of the latent heat ignores the phonon contribution, but this can be expected to be relatively small (the difference between the zero point energies of the R and M1 phases has been estimated to be -17 meV per formula unit using shell model calculations \cite{netsianda08}). Furthermore, upon inclusion of spin polarization in the calculations, the total energies for the R and M1 phases are lowered with respect to the respective NM solutions, which implies that the HSE electronic groundstates for both phases are not the ones described in \cite{eyert11}. The agreement between HSE-level theory and experiment should then be revised. 

\begin{figure*}
\includegraphics[width=16cm]{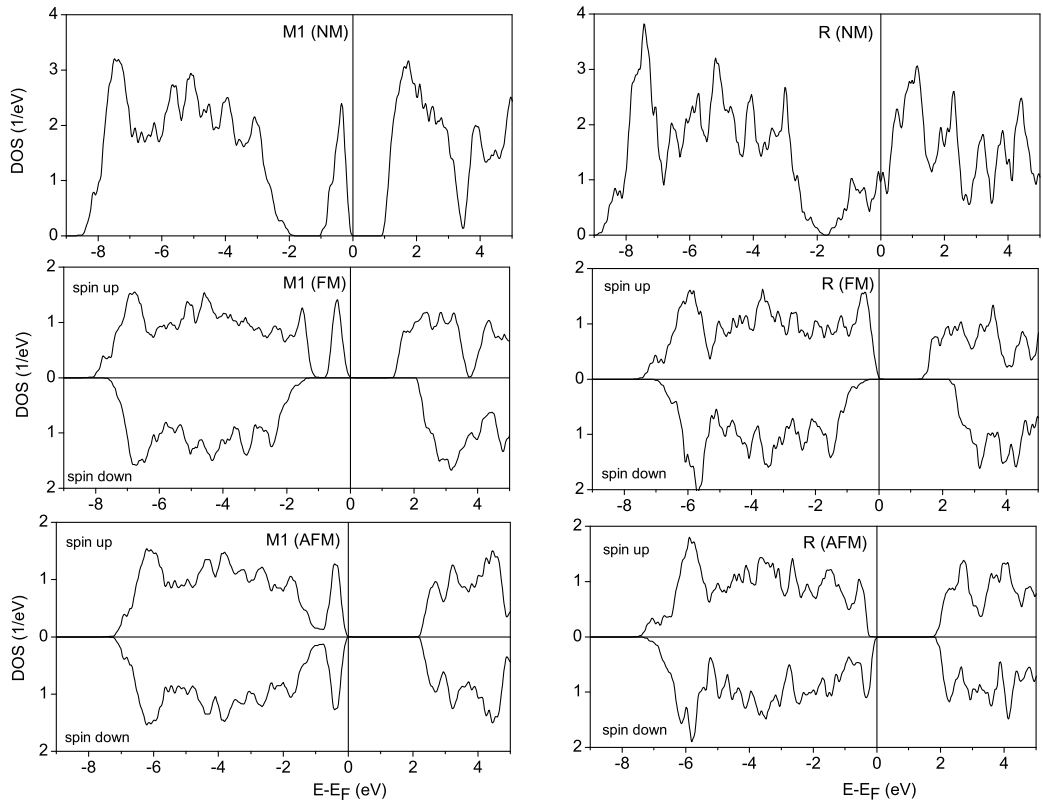}
\caption{\label{fig:two}  Electronic density of states (DOS) of the nonmagnetic (NM), ferromagnetic (FM) and antiferromagnetic (AFM) HSE solutions for the monoclinic (M1) and rutile (R) phases of VO$_2$.}
\end{figure*}

 Allowing spin polarization in the M1 phase calculation, with either antiferromagnetic (AFM, with magnetic moments alternating orientations along the rutile \emph{c} axis) or ferromagnetic (FM) configurations, lowers the total energy by 463 meV and 365 meV per formula unit, respectively, in comparison with the NM calculation. Magnetic moments of $\sim$1 $\mu_B$ are found by integrating the spin density around the V ions, both in the AFM and in the FM case. As shown in Table  \ref{tab:one} and illustrated in the density of states (DOS) plots of Fig. \ref{fig:two}, these lower-energy magnetic solutions have wider band gaps than the NM solution, thus worsening the agreement with experiment. More importantly, the existence of a groundstate with local magnetic moments is in conflict with the well-established non-magnetic character of the M1 phase. Experimentally, VO$_2$(M1) exhibits only a very small, positive and temperature-independent magnetic susceptibility, which is associated with van Vleck paramagnetism, and therefore has no magnetic moments in the groundstate \cite{pouget72, mott75}. The wrong groundstate appears in HSE calculations due to the Hartree-Fock exchange mixed in the HSE functional, which is required for opening the band gap in the M1 phase \cite{eyert11}, but at the same time tends to stabilize localized magnetic moments, in this case excessively (although in other systems the same effect can actually lead to better agreement with experiment \cite{hafner08}). In our calculations we are using the experimental crystal structure, but we have checked that the magnetic moments on the V ions remain stable in the HSE solution for the M1 phase upon changes of +/- 1\% in the cell parameters (HSE-optimised lattice parameters typically deviate less than 1\% from experimental values \cite{heyd04}). We find that the difference between the non-magnetic and the  magnetic solution (taken with FM ordering for calculation convenience) increases 13\% when the cell is expanded, and decreases 13\% when the cell is compressed, but the solutions with magnetic moments are always more stable than the non-magnetic ones.

\begin{table}
\caption{\label{tab:one}
Band gaps obtained from HSE calculations of VO$_2$ phases in different magnetic configurations. Experimental values are from \cite{ladd69, mott75}.}
\begin{ruledtabular}
\begin{tabular}{c c c c c}
		& 		&		&$E_g$(eV)	&\\
		& NM		&FM		&AFM		&Exp.\\
\hline
M1		& 0.98	&1.35		&2.23		&0.6-0.8\\
R		& 0		&1.43		&1.82		&0\\
\end{tabular}
\end{ruledtabular}
\end{table}

For the R phase, HSE gives a ferromagnetic (FM) insulator groundstate  (Fig. \ref{fig:two}), with local magnetic moments of $\sim$1 $\mu_B$ on the V atoms, which is 739 meV per formula unit below the NM solution. This solution is also 506 meV \emph{lower} in energy than the NM solution for the M1 phase. The antiferromagnetic (AFM) solution, with V magnetic moments  in alternate orientations along the \emph{c} axis, also has a band gap, and is close in energy to (but less stable than) the FM groundstate. This situation is again clearly unsatisfactory. Not only it is wrong from a thermodynamic point of view that the groundstate for the R phase  is significantly lower in energy than the M1 phase, but also the magnetic insulator groundstate found by HSE for the VO$_2$(R) structure is in conflict with experimental evidence. The problem here is not the magnetic character, as VO$_2$(R) seems to be paramagnetic with a large temperature-dependent susceptibility \cite{mott75}, but the insulating character of the solution. HSE predicts large band gaps for both the FM and AFM solutions (Table \ref{tab:one}). This is a serious problem, because the metallicity of the R phase is well-established experimentally by conductivity measurements (e.g. \cite{bongers65, corr10}). A note of caution should be added here: our spin-polarized calculations of the R phase are based on magnetically ordered cells, while the proximity in energy between the FM and AFM solutions, and also the experimental evidence, suggest that the system should have paramagnetic disorder. It is therefore still possible in principle (although rather unlikely based on the calculated gap values for the ordered configurations) that the HSE can recover the correct metallic solution for the R phase if magnetic disorder could be accounted for, due to broadening of the bands. Unfortunately, this type of calculation at the HSE level is beyond our computing capabilities at the moment.     

We also note that the HSE approximation to the latent heat is pretty bad, regardless of which solutions are taken for the calculation. The difference in energy between the magnetic R phase and the non-magnetic M1 phase is very large and of opposite sign compared to experiment. If we take the (physically wrong) ground-state solutions for both phases, the absolute value of the HSE latent heat becomes much lower, but still has the wrong sign. This is interesting, because previous calculations based on the local density approximation (LDA), which is unable to account for the band gap opening in VO$_2$(M1), did show good agreement (within 10 meV) with experiment in the latent heat of the transition \cite{wentzcovitch94}. The HSE functional thus performs worse than the simple LDA functional in the description of the relative energies of the phases. 

The existence of incorrect groundstates severely limits the usefulness of the HSE approach in the investigation of this important oxide and its phase transitions. Although it can be argued that in the case of the M1 phase one can still obtain a meaningful solution by forcing a non-spin-polarized calculation, such an approach is not satisfactory. To illustrate why, we consider the case of the tungsten-doped VO$_2$(M1) phase, which is interesting for applications (2 at.\% doping with tungsten can lower the semiconductor-to-metal transition point to room temperature \cite{manning04}). We have investigated the doped oxide using a 96-atom supercell with one V substituted by W. Since the substitution leads to an odd number of electrons per cell, a non-spin polarized calculation necessarily leads to a metallic solution, and in fact this will be the case regardless of the W concentration employed in the simulation. This is an artifact of the spin-restricted calculation; experimental observations confirm that W-doped VO$_2$  remains a semiconductor at low temperatures \cite{manning04, kim07}. In order to deal  correctly with the odd number of electrons, spin polarized calculations are necessary; however, they lead to localized magnetic moments in all the V ions in the cell, i.e. to the wrong groundstate. This was the case even when the initial magnetic moments of the V ions were set to zero in the calculations. Therefore, the problem of W doping in VO$_2$(M1) becomes intractable within the HSE approximation.    

In summary, although the HSE description of the band gap opening in VO$_2$(M1) reported recently \cite{eyert11} is welcomed progress, the results presented here show that the HSE functional does fail in the description of both the electronic structure and the energetics of the transition of VO$_2$. It gives a magnetic groundstate for the M1 phase, a non-metallic groundstate for the R phase, and an R-M1 energy difference in significant disagreement with the experimental latent heat. Vanadium dioxide thus continues to be a challenge to band theory. Despite its succesful record, the HSE functional needs to be used cautiously, particularly in the simulation of the magnetic properties of transition metal oxides. 

We thank Dr Eyert for useful discussion. R.G.C thanks the EPSRC for funding (EP/J001775/1) and for access to HECToR supercomputer via the Materials Chemistry Consortium (EP/F067496). 

\bibliography{vo2}
\end{document}